# Bug Tracking and Reporting System

A.S.Syed Fiaz, N.Devi, S.Aarthi

*Abstract*— This is the world of information. The ever-growing field Information Technology has its many advanced notable features which made it what it was now today. In this world, the information has to be processed, clearly distributed and must be efficiently reachable to the end users intended for that. Otherwise we know it lead to disastrous situations. The other coin of the same phase is it is absolutely necessary to know any bugs that are hither-to faced by the end users. The project "Bug Tracking and Reporting System" aims to provide the solution for that. The Bug Tracker can be made from any two types. The first one being the system side, the other being the services side. Our project deals with the second one.

The paper is wholly dedicated to tracking the bugs that are hither-by arise. The administrator maintains the master details regarding to the bugs id , bugs type, bugs description, bugs severity, bugs status, user details. The administrator too has the authority to update the master details of severity level , status level, etc, modules of the paper. The administrator adds the users and assign them responsibility of completing the paper. Finally on analyzing the paper assigned to the particular user, the administrator can track the bugs, and it is automatically added to the tables containing the bugs , by order of severity and status.

The administrator can know the information in tact the various paper's assigned to various users, their bug tracking status, their description etc in the form of reports from time to time. The paper wholly uses the secure way of tracking the system by implementing and incorporating the Server side scripting. The administrator can now add the project modules, project descriptions etc. He too adds the severity level, its status etc.

The whole beauty of the paper is its high-level and user-friendly interface which mean that is the well based Bug Tracker which helps in tracking the whole system by providing the efficient reporting system. The Bug Tracker can be further by analyzed and further relevant and quick decisions can be taken.

## I. INTRODUCTION

This is the world of information. Bug and issue tracking systems are often implemented as a part of integrated project management system. This approach allows including bug tracking and fixing in a general product development process, fixing bugs in several product versions, automatic generation of a product knowledge base and release notes.

Some bug trackers are designed to be used with distributed revision control software. These distributed bug trackers allow bug reports to be conveniently read, added to the database or updated while a developer is offline. Distributed bug trackers include Fossil. Recently, commercial bug tracking systems have also begun to integrate with distributed version control. Fog Bugz, for example, enables this functionality via the source-control tool, Kiln. Although wikis

Manuscript received Feb 02, 2013.
**A.S.Syed Fiaz**, Department of Computer Science, Anna University, Sona College of Technology, Salem, India.
**N.Devi**, Department of Computer Applications, Periyar University, Muthayammal College of Arts & Science, Namakkal, India.
**S.Aarthi**, Department of Computer Applications, Periyar University, Muthayammal College of Arts & Science, Namakkal, India.

and bug tracking systems are conventionally viewed as distinct types of software, ikiwiki can also be used as a distributed bug tracker. It can manage documents and code as well, in an integrated distributed manner. However, its query functionality is not as advanced or as user-friendly as some other, non-distributed bug trackers such as Bugzilla. Similar statements can be made about org-mode, although it is not wiki software as such.

### A. Defining the problem

The problem in the older system can be defined as the whole project maintenance, users maintenance and their assignment has to be maintained manually. The Software development companies have to face a lot of problems while maintaining manually all the maintenance of the projects, their bugs and their status. This type of problem makes the whole system an inefficient one and thus making a poor and unorganized working. In order to remove this type of problem, So that the paper is planned to develop.

Bug tracking software is a "Defect Tracking System" or a set of scripts which maintain a database of problem reports. Bug tracking software allows individuals or groups of developers to keep track of outstanding bugs in the product effectively. Bug tracking software can track bugs and changes, communicate with members, submit and review patches, and manage quality assurance.

This web-based business application is a great tool for assigning and tracking issues and tasks during software development and any other projects that involve teams of two or more people.

## II. ADMIN MODULE

The administrator too has the authority to update the master details of severity level , status level, etc, modules of the project. The administrator adds the users and assign them responsibility of completing the project. Finally on analyzing the project assigned to the particular user, the administrator can track the bugs, and it is automatically added to the tables containing the bugs , by order of severity and status.

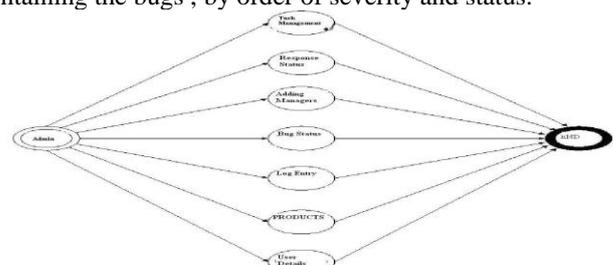

Fig (a). Admin module in Bug Tracking System

# Bug Tracking and Reporting System

## III. MANAGER MODULE

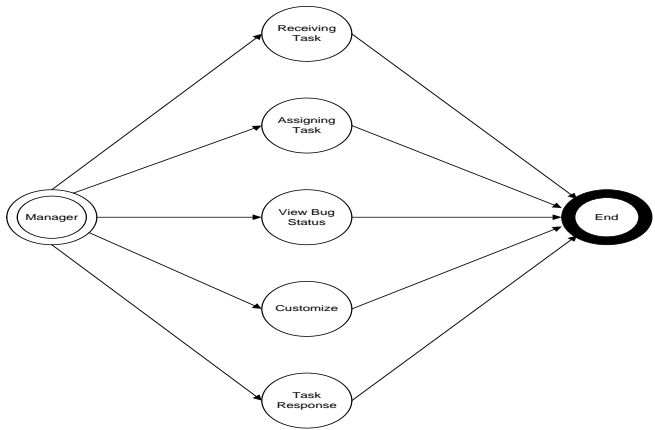

Fig (b). *Manager module in Bug Tracking System*

The administrator can know the information in tact the various projects assigned to various users, their bug tracking status, their description etc in the form of reports from time to time. The project wholly uses the secure way of tracking the system by implementing and incorporating the Server side scripting. The administrator can now add the project modules, project descriptions etc. He too adds the severity level, its status etc.

## IV. DEVELOPER MODULE

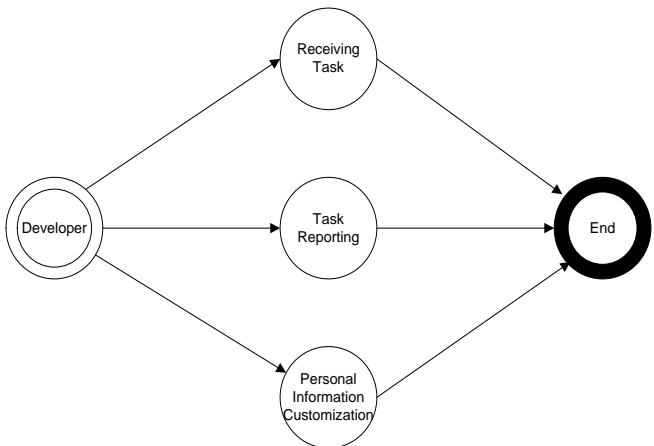

Fig (c). Developer Module in Bug Tracking System

## V. STRUCTURE FLOW DIAGRAM

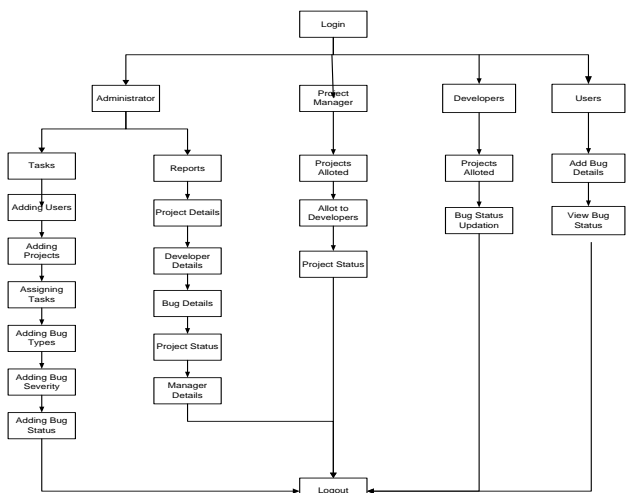

Fig (d). *Structure of Bug Tracking System*

## VI. TABLE DESIGN

### A. Project bug

| Field name | Data type | Description |
|---|---|---|
| Bugid | Number | Not Null |
| Bugname | Text | Not Null |
| Proid | Number | Not Null |
| Proname | Text | Not Null |
| Staid | Number | Not Null |
| Sevid | Number | Not Null |

### B. Project status

| Field name | Data type | Description |
|---|---|---|
| Username | Text | Not Null |
| projname | Text | Not Null |
| projstatus | Text | Not Null |

### C. Bug types

| Field name | Data type | Description |
|---|---|---|
| typid | Number | Primary key |
| typname | Text | Not Null |
| typdesc | Text | Not Null |

The paper "Bug Tracking and Reporting System" helps the Software Development companies to track the exact status of different project bugs and to rectify their errors in time and in right manner.

## VII. DEVELOPING SOLUTION STRATEGIES

The main objective of the proposed system is to full analyze the bugs and report the same to the administrator in an efficient manner so that he can get right information at right times. The paper objective is to fully systemize everything so that the possibilities of bugs should be reduced at all levels.

### A. Benifits

1) To track the status level of each project.
2) To track the status level of each bug in the project module.
3) To assign the projects to the users by the administrator.
4) To add the bugs by the administrator.
5) To add the status, severity levels by the administrator
6) To add a detailed bug information.
7) To add the modules in the project and to track the person developing it.
8) To add the project status levels by the project managers
9) To add the project bug levels by the users.
10) To give an efficient reporting system so that right decisions can be taken and at right times.
11) Moreover to make the system fully utilize to reduce the bugs.

Bug Tracking and Reporting System## VIII. IMPLEMENTATON PROCEDURE

Implementation is the stage, which is crucial in the life cycle of the new system designed. The main stage in the implementation is planning, training, system testing. Implementation is converting a new or revised system into an operational one. Conversion is the main aspect of implementation. It is the process of changing from the old system to the new one. After system is implemented, user conducts a review of the system. It is used to gather information for the maintenance of the system. The basic review method is a data collection method of questionnaire, interview etc.

## IX. COST ESTIMATION AND SCHEDULING

Cost Estimation can be made either top-down or bottom-up. Top-down estimate starts with system level costs, work out the costs of computing resources, development staff, configuration management, quality assurance, system integration, training and publications.

Constructive Cost Model (COCOMO) is the top-level model. Basic COCOMO is applicable to large majority of software project.

BASIC COCOMO

| Percentage | Phase | Activity |
|---|---|---|
| 20% | Engineering | Planning, specification of function or performance etc., |
| 20% | Design | Specification of design |
| 17% | Code and unit or module test | Final product is a set of program modules accepted by the test and integration system |
| 43% | System Test and Integration | Full system involvement, hardware, software, operators etc., Final product is a system ready for acceptance test. |

Fig (e). *Basic Cocomo Constructive Model*

The Paper titled "Bug Tracking and Reporting System" adopts the Cost Estimation in a planned and full-fledged manner. It follows the Cost-Benefit analysis while making the Cost Estimation of the implementation of the wireless based communication system.

Input design is the process of converting user-oriented inputs to a computer-based format. The quality of the system input determines the quality of system output. Input design determines the format and validation criteria for data entering to the system.

## X. INPUT OF BUG TRACKING SYSTEM

Input data is a part of the overall system design, which requires very careful attention. If the data going into the system is incorrect then the processing and output will magnify these errors. Input can be categorized as internal, external, operational, computerized and interactive. The analysis phase should consider the impact of the inputs on the system as a whole and on the other systems.

In this paper, the inputs are designed is such a way that occurrence of errors are minimized to its maximum since only authorized user or administrator can able to access this tool. The input is given by the administrators are checked at the entry form itself. So there is no chance of unauthorized accessing of the tool. Any abnormality found in the inputs are checked and handled effectively. Input design features can ensure the reliability of a system and produce results from accurate data or they can result in the production of erroneous information.

## XI. OUTPUT OF BUG TRACKING SYSTEM

Computer output is the most important and direct source of information to the users. Designing the output should proceed in an organized, well thought out manner. The right output must be developed while ensuring that each output element is designed so that people will find easy to use the system. When analyst designs the output, they identify the specific output that is needed to meet the information requirements.

The success and failure of the system depends on the output, though a system looks attractive and user friendly, the output it produces decides upon the usage of the system. The outputs generated by the system are checked for its consistency, and output is provided simple so that user can handle them with ease. For many end user, outputs is the main reason for developing the system and the basis on which they will evaluate the usefulness of the application.

## XII. QUALITY ASSURANCE

Quality Assurance is 'a planned and systematic pattern of all actions necessary to provide adequate confidence that the item or product conform to established technical requirements'. This package is tested for Software Quality Assurance. During analysis and design phase, it is verified whether the requirements are satisfied by the design documents and found that the source code is consistent with the requirements specifications and design documentation. The Acceptance test plan is executed and the developer is satisfied with the expected outcomes.

## XIII. SECURITY TECHNOLOGIES AND POLICIES

Software safety is a software quality assurance activity that focuses on the identification and assessment of potential hazards that may impact software negatively and cause the entire system to fail.

The system is highly secured that all Forms are interlinked and follows the login so that only authorized users who have known the correct password can only enter and work with the system. The package could not exposed to outside hosts.

## XIV. RESULT

The older system is a system which suffers from a lot of disadvantages:
1) Limitations of the older System
2) Has to maintain the whole system of
3) software development manually
4) Has to maintain the details of the project managers manually.

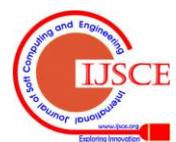

5) Has to maintain the details of the status of different projects manually.
6) Has to maintain the bug details, project descriptions, project status details, project description details, project list details manually.
7) No efficient reporting system.

In this paper we presented the results to overcomes the problem with older system. The bug tracking system fulfills different requirements of administrator and employee of a software development organization efficiently. The specific purpose of the system is to gather and resolve issues that arise in different projects handled by the organization. The advantages of this paper is:

*A. Advantages*

1) The different groups and representatives can interact each through internet.
2) Main objective of the system is to gather and resolve issues (bugs) that arise in different projects of the organization.
3) Reduce The Timing.
4) Internet application.

## XV. CONCLUSION

This Paper Bug Tracking and Reporting System helps an Software Concern to detect and manage the bug in their products effectively-efficiently. Utilizing bug tracking software can assist in troubleshooting errors for testing and for development processes. With the ability to provide comprehensive reports, documentation, searching capabilities, tracking bugs and issues, bug tracking software is a great tool for those software development needs.

Depending on your development needs and the bug tracking software, you can hope to gain several benefits from bug tracking software.

Some of the benefits are:
1) Improve communications between groups of people
2) Increase the quality of the software
3) Improve customer satisfaction with bug free software
4) Provides a form of accountability
5) Increases overall productivity

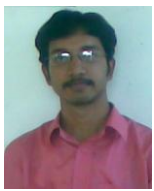
A.S.Syed Fiaz received BE in Computer Science and Engineering from Sona College of Technology, Anna University, Chennai 2010 & Currently he is pursuing his final year ME in Computer Science and Engineering from Sona College of Technology, Anna University, Chennai 2013 . His area of interests are Cloud computing, Computer Networks and Database Management Systems.

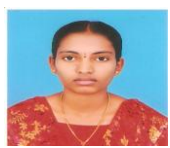
 **N.Devi**, Pursuing her Bachelor in Computer Appliactions, Muthayammal College of Arts & Science, Periyar University, Namakkal, India. Her area of interests are Computer Networking and Database Management Systems.

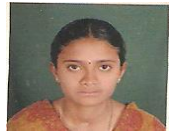
**S.Aarthi,** Pursuing her Bachelor in Computer Appliactions, Muthayammal College of Arts & Science, Periyar University, Namakkal, India. Her area of interests are Wireless Networks and Database Management Systems.